\documentclass[runningheads]{llncs}
\usepackage[T1]{fontenc}
\usepackage[latin9]{inputenc}
\usepackage{caption}
\usepackage{array}
\usepackage{multirow}
\usepackage{amsmath}
\usepackage{amsthm}
\usepackage{stackrel}
\usepackage{graphicx}
\usepackage{algorithm,algpseudocode}
\usepackage{commath}
\usepackage{enumitem}
\usepackage[colorinlistoftodos]{todonotes}
\usepackage{float}
\usepackage{bbold}
\usepackage{mathtools, nccmath}

\begin{document}
\title{HYCEDIS: HYbrid Confidence Engine for Deep Document Intelligence System}

\author{Bao-Sinh Nguyen\inst{1} \and Quang-Bach Tran\inst{1} \and Tuan-Anh D. Nguyen\inst{1}  \and Hung Le\inst{2}}

\authorrunning{Nguyen et al.}
\titlerunning{HYCEDIS}

\institute{Cinnamon AI \\10th floor, Geleximco building, 36 Hoang Cau, Dong Da, Hanoi, Vietnam. \\ \email{\{simon, neath, tadashi\}@cinnamon.is} \and
Deakin University, Australia.\\
\email{thai.le@deakin.edu.au}}

\maketitle
\begin{abstract}
Measuring the confidence of AI models is critical for safely deploying AI in real-world industrial systems. One important application of confidence measurement is information extraction from scanned documents. However, there exists no solution to provide reliable confidence score for current state-of-the-art deep-learning-based information extractors. In this paper, we propose a complete and novel architecture to measure confidence of current deep learning models in document information extraction task. Our architecture consists of a Multi-modal Conformal Predictor and a Variational Cluster-oriented Anomaly Detector, trained to faithfully estimate  its confidence on its outputs without the need of host models modification. We evaluate our architecture on real-wold datasets, not only outperforming competing confidence estimators by a huge margin but also demonstrating generalization ability  to out-of-distribution data. 
\end{abstract}


\section{Introduction}
Recent advances in machine learning enables creations of automatic information extractors that can read the input document in image format, locate and understand relevant text lines before organizing the information into computer-readable format for further analysis \cite{yang2017learning,he2018end}.
Despite these successes, in critical domains such as healthcare and banking, humans still have to involve to scrutinize AI outputs as there is no room for AI errors in making important decisions that can affect human life.
Confidence score estimation is one critical step towards implementing practical industrial systems wherein AI automates most of the operations yet human will intervene if necessary \cite{zheng2017hybrid}. 

Unfortunately, to the best of our knowledge, there exists no holistic solution to reliably estimate the confidence score for the task of document information extraction. Current confidence score approaches are either generic methods verified only for simple image classification tasks \cite{gal2016dropout} or applied only for part of the information extraction process \cite{mor2018confidence}.

In this paper, we introduce a novel neural architecture that can judge the result of extracted structured information from documents provided by the information extracting neural networks (hereafter referred to as the IE Networks). Our architecture is hybrid, consisting of two models, which are a Multi-modal Conformal Predictor (MCP) and an Variational Cluster-oriented Anomaly Detector (VCAD). The former aims to combine the neural signals from 3 main stages of information extraction processes including text-box localization, OCR, and key-value recognition to predict the confidence level for each extracted key-value output. The later computes anomaly scores for the raw input document image, providing the MCP with additional features to produce better confidence estimation. The VCAD works on global, low-level features and plays a critical role in lifting the burden of detecting outliers off the MCP, which focuses more on local, high-level features.

We demonstrate the capacity of our proposed architecture on  real-world invoice datasets: SROIE \cite{huang2019icdar2019}, CORD \cite{park2019cord}) and 2 in-house datasets. The experimental results demonstrate that our method outperforms various confidence estimator baselines (including Droupout \cite{gal2016dropout}, temperature scaling \cite{guo2017calibration}). In short, we summarize our contribution as follows:
\begin{itemize}
    \item We propose a \textit{Multi-modal Conformal Predictor (MCP)} using a Feature Fusion module over 3 Feature Encoders to fuse signals extracted from IE Networks and compute the confidence score of the IE Networks' outputs.
    
    \item We provide a \textit{Variational Cluster-oriented Anomaly Detector (VCAD)} to equip the MCP with an ability to handle out-of-distribution data.
    
    \item We unify the proposed MCP and VCAD in a single hybrid confidence engine, dubbed as HYCEDIS, that for the first time, can well estimate the confidence of document intelligent system. 
      
    \item We conduct intensive experiments on 4 datasets with detailed ablation studies to show the effectiveness and generalization of our hybrid architecture on real-world problems.
\end{itemize}
\section{Background\label{sec:bg}}
A typical Document Intelligence System consists of multiple smaller steps: text detection, text recognition and information extraction (IE).
Given a document image, the usual first step is to detect text lines, using segmentation \cite{baek2019character,he2017single,long2018textsnake}  or object detection method \cite{liao2017textboxes,liao2018rotation,liu2017deep}. The detected text line images can each go through an OCR model to transcribe into text \cite{graves2006connectionist}. After all text contents are transcribed, the relevant text entities can be extracted, using entity recognition (sequence tagging) method \cite{yao2019graph,liu2019graph,xu2020layoutlm}, segmentation-based method \cite{yang2017learning,dang2019end}, or graph-based method \cite{Qian2018,Liu2019,Vedova2019} which formulates the document layout as a graph of text-lines/words.

In this paper, we adopt a common IE Network that consisted of 3 main modules: text detection (Layout Analysis), text recognition (CRNN) and graph-based information extraction model (Graph KV). The text detection model shares the same architecture with \cite{baek2019character} which utilizes segmentation masks to detect text-lines in the document image. The text recognition (CRNN) uses popular CNN+Bi-LSTM+CTC-loss architecture to transcribe each text-line images into text. Finally, the GCN model \cite{Liu2019} performs the node classification tasks from the input document graph constructed from the text-lines' location and text to extract relevant information. Here, for our problems, we classify each node into different key types that represent the categories of the text-line.

\section{Methodology}
\begin{figure*}[t]
    \centering
    \includegraphics[width=0.99\linewidth,height=0.35\textheight]{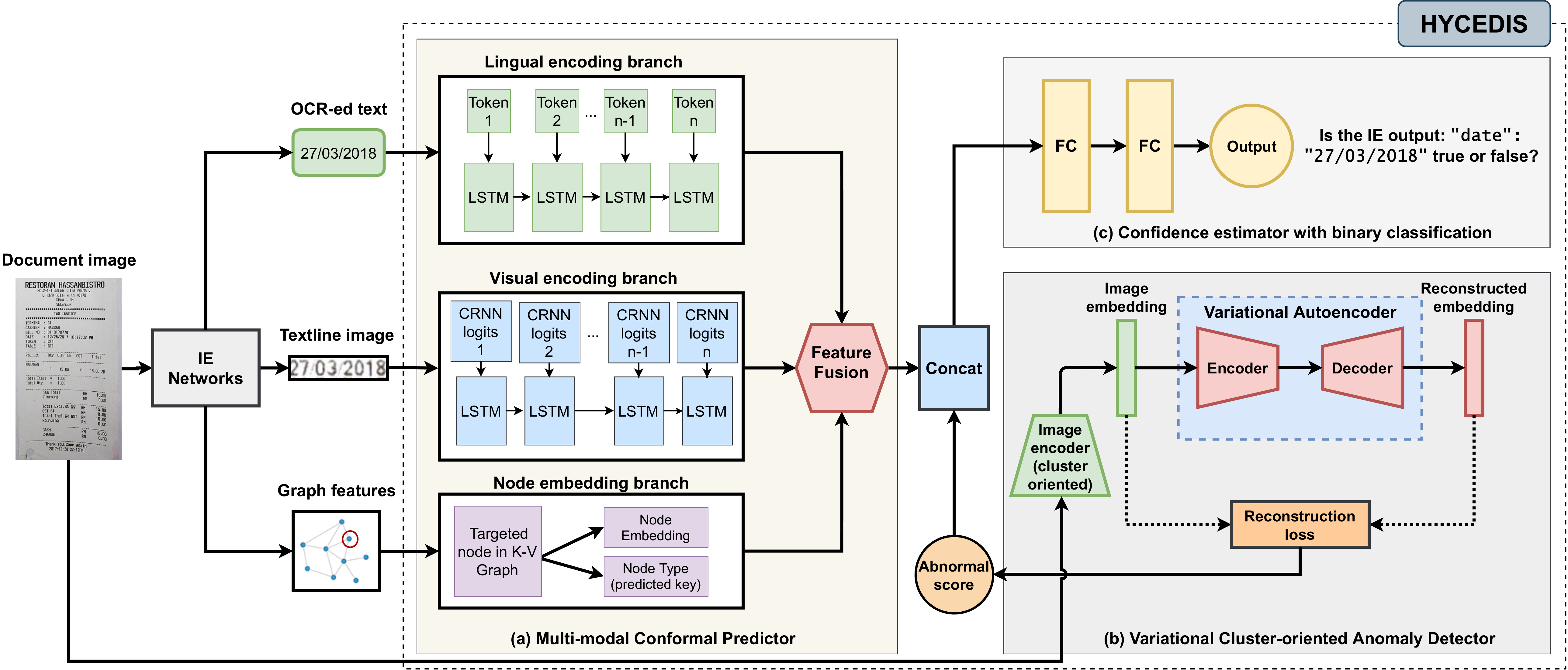}
    \caption{\small{HYCEDIS architecture. (a) The Multi-modal Conformal Predictor (MCP). (b) The Variational Cluster-oriented Anomaly Detector (VCAD). (c) Confidence estimator (CE). MCP's output vector, plus the VCAD's abnormal score, is fed to fully-connected layers to produce the final output of HYCEDIS, indicating whether the extracted field true or false.}}
    \label{fig:architecture_CP}
\end{figure*}

\subsection{Multi-modal Conformal Predictor (MCP)}\label{subsec:mcp}

Given extracted intermediate features of IE Networks, our Multi-modal Conformal Predictor aims to estimate the confidence score through predicting whether the final output is \textit{true} or \textit{false}. The MCP architecture (see Figure \ref{fig:architecture_CP}(a)) contains two main components which are Feature Encoding and Feature Fusion. The Feature Encoding extracts features from different layers of trained IE Networks while the Feature Fusion combine them for predicting the final output.

\paragraph{Feature Encoding}
Motivated by designs of late-fusion multi-view sequential learning approaches \cite{chung2017lip}, three components of the Feature Encoding layers are independent processing streams including visual, lingual and structural feature encoders. In particular, the visual feature encoder is a many-to-one LSTM $f_{VF}(\cdot)$ that captures the visual information embedded in the CRNN of the IE Network. It takes the CRNN's logits (whose shape is $T \times F_{in\_vis}$ where $T$ is the number of timesteps) as input and outputs a vector of size $F_{out\_vis}$, which represents the knowledge of the IE Networks on its OCR model's neural activations given the input image. In particular, for the $i$-th extracted text-line image $I_i$, we compute it as: $E^{vis}_i = f_{VF}(CRNN(I_i))$.
   
The lingual feature encoder, which is also implemented as a many-to-one LSTM  $f_{LF}(\cdot)$, processes the predicted OCR texts of the IE Networks. Each OCR-ed character in the text is represented as an one-hot vector with the size that equals to the size of the corpus. For the $i$-th extracted OCR-ed text, the LSTM takes a sequence of these one-hot vectors (denoted by $text_i$, whose shape is $T\times F_{in\_OCR}{}$) and produces an output vector of size $F_{out\_OCR}$, representing the knowledge of the IE Networks on the linguistic meaning and the syntactic pattern of its OCR-ed outputs. We compute it as: $E^{OCR}_i = f_{LF}(text_i)$.

The structural feature encoding $f_{SF}(\cdot)$ is a feed-forward neural network that accesses the information from the final layer of the IE Networks  (node classification) -- the Graph KV module. Here, the logits before softmax layer of the $i$-th node in the graph (corresponding to the $i$-th text box extracted from the document), denoted by ${logit}^{(KV)}_i$, is the input of the structural feature encoding, and the corresponding output is node embedding vector $E^{node}_i$ representing the knowledge of the IE networks on its final decision (node classification): $E^{node}_i = f_{SF}({logit}^{(KV)}_i)$.

\paragraph{Feature Fusion}
The Feature Fusion network $f_{Fusion}$ takes the three outputs from the Feature Encoding module and produces the ultimate feature vector. We use simple concatenation and Bi-linear pooling \cite{yu2017multi} as two options for Feature Fusion. Bi-linear pooling use outer-product to combine inputs of different modalities. For simple concatenation, we just concatenate three vectors. For Bi-linear Pooling, we first pool the pair of $E_i^{vis}$ and $E_i^{OCR}$, and then pool the resulting vector with $E_i^{node}$ to get the pooled output $F_i$:

\begin{equation}
    F_{i} = f_{Fusion}(E_i^{vis}, E_i^{OCR}, E_i^{node})
\end{equation}

\subsection{Variational Cluster-oriented Anomaly Detector (VCAD)}

The anomaly detector aims to detect which input image is normal or abnormal, thus bolsters the MCP by a measurement of the normality that the input has. Specifically, the input to the anomaly detector is a compressed representation of the document image, and the output is a score in the range $[0, 1]$ indicating the level of anomaly of the input. This score serves as an additional input to the confidence estimator.

\paragraph{Representing image data with cluster-oriented embeddings}

In this section, we describe the representation learning of document images. Firstly, the training dataset was classified into some categories based on the appearance and the layout structure of the document image. Then we train a CNN-based image encoder to map each document image into a lower-dimensional vector representation. Here, the CNN architecture is MobileNet \cite{sandler2018mobilenetv2}. We adopt the triplet loss \cite{huang2012bounding} to learn the compressed representation, wherein the embeddings of images from the same category tend to form a cluster in the embedding space.

\paragraph{Anomaly detector training}
After constructing embeddings for training images, we build a Variational Auto Encoder (VAE) \cite{kingma2013auto,rezende2014stochastic} as our anomaly detector (Figure \ref{fig:architecture_CP}(b)).
The VAE outlier detector is first trained on a set of normal (inlier) data to reconstruct the input it receives, with the standard VAE loss function which is the sum of KL term and reconstruction loss: 

\begin{equation}
    \mathcal{L_\mathrm{VAE}}(x;\theta,\phi) = -KL(q_\phi(z|x)||p_\theta(z)) + \frac{1}{L}\sum_{l=1}^{L}\log{p_\theta(x|z^{(l)})}
\end{equation}
where $x$, $z$ and $L$ denote the VAE's input, latent variable, and number of samples, respectively. $q_\phi$ represents the encoder and $p_\theta$ the decoder of VAE.

If the input data cannot be reconstructed well, the reconstruction error (implemented as L1 loss between VAE's input and output) is high and the data can be flagged as an outlier. We apply the min-max normalization \cite{akcay2018ganomaly} to the reconstruction losses in order to get the corresponding abnormal scores in the range of $[0, 1]$.

\subsection{Hybrid confidence estimation}\label{subsec:hce}
After getting the scalar output from our VCAD, we simply concatenate this scalar with the output of the Feature Fusion module in the MCP. The resulting vector is fed to a confidence estimator (CE), which is implemented as a 2-layer feed-forward neural network. We freeze the VCAD and train the CE and MCP end-to-end on the training data as the set of IE's predictions in the training dataset.

In particular, let $x_i$ denote the input document image, the function $IE(\cdot)$ denote our pipeline of IE networks. The output of IE system is $\hat{v}_i = IE(x_{i})$. More specifically, $\hat{v}_i = \{\hat{v}_{ik}\}_{k=1:K_i}$ is the set of $K_i$ predictions where each $\hat{v}_{ik}$ contains location information along with extracted text corresponding to a particular key (e.g: \texttt{\textit{\{'location': [123,234,184,246], 'text': '27/03/2018', 'key': 'date'\}}}). We also have the ground truth $v_i = \{v_{ij}\}_{j=1:J_i}$ is the set of $J_i$ elements presented in the $i\mathrm{-th}$ document.

Let $F_{ik}$ denote the input of our CE corresponding to prediction $\hat{v}_{ik}$, the CE is represented by the function $f_{CE}(\cdot)$ yielding the softmax output $p_{ik} = f_{CE}(F_{ik})$. The label for confidence estimation task is 

\begin{equation}
y_{ik} = \mathbb{1} \{\exists j \in \{1:J_i\}\mid v_{ij} \overset{\mathrm{match}}{=} \hat{v}_{ik}\}    
\end{equation}

The IE's output is considered to match the ground truth element if both the text contents and the keys match and the locations' IoU is greater than a threshold ( $0.3$ in this paper). $y_{ik}$ is $1$ if IE's prediction matches a ground truth element (be correct) and vice versa. Then the loss function is the standard binary cross-entropy loss with label $y_{ik}$ and probability $p_{ik}$.
\section{Experiments}
\subsection{Datasets and evaluation metrics} \label{Evaluation metrics}

\subsubsection{Datasets}
We collect 4 Invoice-like datasets and divide them into 2 tasks, corresponding to English and Japanese language used in the data. For each task, we use the bigger dataset as the main one, and the smaller as the out-of-distribution (OOD) dataset with respect to the main dataset. 

We first use pre-trained  IE Networks (see Sec. \ref{sec:bg}) to generate the intermediate features for the MCP as mentioned in Sec. \ref{subsec:mcp}. The outputs of the IE Networks and the ground-truth IE outputs are used to produce labels for the confidence estimation task (Sec. \ref{subsec:hce}). 

We only train the confidence models on the training dataset and benchmark them on the testing and corresponding OOD datasets. The evaluation on OOD data is a challenging benchmark since the OOD dataset is totally different from the main one in terms of layout, background and writing styles. Moreover, since the OOD datasets can have different type of keys from those in the main one, we only test the models on fields that share common keys with the main dataset.

\begin{enumerate}[label=(\alph*), topsep=0pt]

\item Public datasets (\textit{English})

\textit{SROIE - Main dataset.} SROIE \cite{huang2019icdar2019} is a dataset of scanned receipts. There are 4 keys: \textit{address}, \textit{company}, \textit{date}, \textit{total}. The training set has 626 files corresponding to 3859 IE's output key-value fields. We further hold 10\% of the training as the validation set.  The statistics for the test set are 341 files and 1,640 fields, respectively. 

\textit{CORD - OOD dataset.} CORD \cite{park2019cord} contains receipts collected from Indonesian shops and restaurants. Compared to SROIE, CORD document images are captured in the wild, thus the data is noisy and low in quality. CORD field shares only one key with SROIE, which is \textit{total}. We use the CORD-dev set which contains 100 files correspoding to 103 IE's output fields. 

\item In-house datasets (\textit{Japanese})

\textit{In-house 1 - Main dataset.} In-house 1 is a dataset containing Japanese invoice documents collected from several vendors. There are 25 keys. Example keys are \textit{issued\_date, total\_amount, tax, item\_name, item\_amount}. The training set has 835 files corresponding to 24,697 IE's output fields, and the test set has 338 files and 10,898 fields.

\textit{In-house 2 - OOD dataset.} In-house 2 consists of 68 invoice documents from another Japanese company. The document pattern is quite different to the In-house 1 dataset. The two in-house dataset share  4 key types in common, resulting in 3,887 IE's output fields.
\end{enumerate}

\subsubsection{Evaluation metrics}
We use the popular \textit{Area Under the Receiver Operating Characteristic Curve} (\textit{AUC}) \cite{mor2018confidence,ayhan2018test,mandelbaum2017distance,hein2019relu} and 
\textit{Expected Calibration Error} (\textit{ECE}) \cite{naeini2015obtaining} metrics for measuring the performance of confidence predictors.



\subsection{Experimental baselines}


\textbf{Softmax Threshold.}
Our IE pipeline consists of multiple sequential models, so we adapted \cite{hendrycks2016baseline} by combining both softmax probabilities from OCR and KV models using multiplication (i.e: \({p_{final}} = {p_{OCR}} * {p_{KV}}\)). We then specify a threshold score and considered examples with higher-than-threshold softmax probability as correctly predicted one, and vice versa. The threshold score is tuned on the training dataset.

\noindent\textbf{Temperature Scaling.}
Temperature scaling \cite{guo2017calibration} is a technique that post-processes the neural networks to make them calibrated in term of confidence. Temperature scaling divides the logits (inputs to the softmax function) by a learned scalar parameter $T$ (temperature). We learn this parameter on a validation set, where $T$ is chosen to minimize negative log-likelihood.

\noindent\textbf{Softmax Classifier.}
Instead of only utilizing the softmax probability of the predicted class as  Softmax Threshold, Softmax Classifier is more advanced by making use of the whole softmax vector. Particularly, we build a simple classifier using a feed-forward neural network. The input for the network is the concatenation of the OCR model's softmax vector and the KV model's one.

\noindent\textbf{Monte Carlo Dropout.}
MC Dropout \cite{gal2016dropout} belongs to the class of Bayesian/variational approaches. By keeping the dropout enabled at test time, we can obtain the variance of the neural network's outputs, and this variance indicates the level of uncertainty. We apply MC Dropout on our KV model, which is the final model in the pipeline. 
\enlargethispage{1\baselineskip}
\subsection{Benchmarking results}

\subsubsection{Ablation study} We ablate the effect of VCAD and MCP on the whole hybrid system. Table \ref{tab:result_of_proposed_models} reports the results on SROIE dataset. Without VCAD, the proposed model achieves best AUC score of 86.90\% using bi-linear pooling fusion strategy. 
Simpler concatenation method underperforms by about 3\% demonstrating the importance of using outer-product to retain bit-level relationships among 3 modalities.  
When the VCAD is integrated, it consistently improves the performance of all fusion methods. Hence, the full hybrid HYCEDIS architecture can reach 88.12\% AUC. Similar behaviors can be found with measurement using ECE metric. 

\begin{table}[]
\caption{Ablation study on SROIE dataset}
\label{tab:result_of_proposed_models}
\centering
\begin{tabular}{l|r|r}
\hline
\multicolumn{1}{c|}{Methods} & \multicolumn{1}{l|}{ECE} & \multicolumn{1}{l}{AUC} \\ \hline 
MCP (concatenation)            & 0.1525                   & 83.75                  \\ 
MCP (bilinear pooling)                   & 0.1175                   & 86.90                  \\ 
MCP (concatenation) + VCAD          & 0.1385                   & 84.37                  \\ 
MCP (bilinear pooling)  + VCAD                   & \textbf{0.1002}          & \textbf{88.12}         \\ \hline
\end{tabular}
\end{table}

\begin{table}[ht]
\caption{Performance comparison of baselines and proposed methods on SROIE and CORD datasets}
\label{tab:result_public_dataset}
\centering
\begin{tabular}{l|c|c|c|c}
\hline
\multirow{2}{*}{Methods} & \multicolumn{2}{c|}{SROIE} & \multicolumn{2}{c}{CORD} \\ \cline{2-5} 
                         & ECE          & AUC         & ECE         & AUC         \\ \hline
                         
Softmax threshold             & 0.1525          & 83.75           & 0.1731          & 66.91          \\ 
Softmax classifier            & 0.1400          & 85.50          & 0.3289          & 54.91          \\ 
MC Dropout                    & 0.1175          & 86.90          & 0.5446          & 43.52          \\ 
Temperature scaling           & 0.1385          & 84.37          & 0.3787          & 74.58          \\ 
MCP                           & 0.1124 & 86.40 & 0.1432 & 75.12 \\ 
HYCEDIS                     & \textbf{0.1002} & \textbf{88.12} & \textbf{0.1259} & \textbf{77.45} \\ \hline
\end{tabular}
\end{table}

\begin{table}[ht]
\caption{Performance comparison of baselines and proposed methods on In-house datasets}
\label{tab:final_result_in_house}
\centering
\begin{tabular}{l|c|c|c|c}
\hline
\multirow{2}{*}{Methods} & \multicolumn{2}{c|}{In-house 1} & \multicolumn{2}{c}{In-house 2} \\ \cline{2-5} 
                         & ECE          & AUC         & ECE         & AUC         \\ \hline
                         
Softmax threshold             & \multicolumn{1}{r|}{0.1285} & 68.79                               & \multicolumn{1}{r|}{0.5885} & 53.38                               \\ 
Softmax classifier            & \multicolumn{1}{r|}{0.2810} & 71.43                               & \multicolumn{1}{r|}{0.3945} & 51.22                               \\ 
MC Dropout                    & \multicolumn{1}{r|}{0.3733} & 66.14                               & \multicolumn{1}{r|}{0.3621}  & 48.20                               \\ 
Temperature scaling           & \multicolumn{1}{r|}{0.1728} & 64.00                               & \multicolumn{1}{r|}{0.5879} & 58.18                               \\ 
MCP                           & 0.0782             & \multicolumn{1}{c|}{{86.32}} & 0.3348             & \multicolumn{1}{c}{60.12} \\ 
HYCEDIS                    & \textbf{0.0712}             & \multicolumn{1}{c|}{\textbf{90.12}} & \textbf{0.3019}             & \multicolumn{1}{c}{\textbf{61.90}} \\ \hline
\end{tabular}
\end{table}


\subsubsection{Public English datasets result} Table \ref{tab:result_public_dataset} shows the performance of all models on public datasets. On both SROIE and its OOD CORD dataset, our full HYCEDIS is consistently the best performer regarding both ECE and AUC scores. Our MCP is the runner-up under ECE metric. The improvements of MCP in AUC and ECE suggests that the signals from intermediate features extracted from text-line images, OCR-ed text and graph structure help improve the accuracy of the softmax-based methods which only rely on some softmax layers of the IE Networks. In addition, when combined with VCAD, the AUC score is further increased and the ECE also downgrades. That manifests the contribution of our VCAD model. We can see a significant performance drop from baselines such as MC-Dropout when being tested on OOD CORD data. Our methods alleviate this issue, maintaining a moderate generalization to strange data.

\subsubsection{In-house Japanese datasets result} We also benchmark the models on two in-house datasets. In Table \ref{tab:final_result_in_house}, our model continues to show the superior performance compared with other baselines. Our MCP model improves about 14.89\% and 2\% AUC score and reduces 0.0503 and 0.0273 ECE score in In-house 1 and In-house 2 datasets, respectively. When adding VCAD, the performance is improved around 3.82\% on In-house 1 dataset and 2.78\% on In-house dataset, which again validates our hypothesis on using anomaly detector to enhance conformal predictor. 
\section{Conclusion}
\enlargethispage{2\baselineskip}
We have introduced a holistic confidence score architecture that aims to verify the result of IE Networks in document understanding tasks. Our architecture takes advantages of a Multi-modal Conformal Predictor and a Variational Cluster-oriented Anomaly Detector to predict whether the IE Networks' output correct or not using features of different granularity. Our hybrid approach surpasses prior confidence estimation methods by a huge margin in benchmarks with invoice datasets. Remarkably, it demonstrates a capability of generalization to out-of-distribution datasets.
\bibliography{ref}
\bibliographystyle{splncs04}

\end{document}